\documentstyle[12pt]{article}
\input epsf
\catcode`\@=11
\long\def\@makefntext#1{
\protect\noindent \hbox to 3.2pt {\hskip-.9pt
$^{{\ninerm\@thefnmark}}$\hfil}#1\hfill}
\def\@makefnmark{\hbox to 0pt{$^{\@thefnmark}$\hss}}
\def\ps@myheadings{\let\@mkboth\@gobbletwo
\def\@oddhead{\hbox{}
\rightmark\hfil\ninerm\thepage}
\def\@oddfoot{}\def\@evenhead{\ninerm\thepage\hfil
\leftmark\hbox{}}\def\@evenfoot{}
\def\sectionmark##1{}\def\subsectionmark##1{}}

\renewcommand{\thefootnote}{\fnsymbol{footnote}}

\newcounter{sectionc}\newcounter{subsectionc}\newcounter{subsubsectionc}
\renewcommand{\section}[1] {\vspace*{0.6cm}\addtocounter{sectionc}{1}
\setcounter{subsectionc}{0}\setcounter{subsubsectionc}{0}\noindent
	{\normalsize\bf\thesectionc. #1}\par\vspace*{0.4cm}}
\renewcommand{\subsection}[1]
{\vspace*{0.6cm}\addtocounter{subsectionc}{1}
	\setcounter{subsubsectionc}{0}\noindent
	{\normalsize\it\thesectionc.\thesubsectionc. #1}\par\vspace*{0.4cm}}
\renewcommand{\subsubsection}[1]
{\vspace*{0.6cm}\addtocounter{subsubsectionc}{1}
	\noindent
{\normalsize\rm\thesectionc.\thesubsectionc.\thesubsubsectionc.
	#1}\par\vspace*{0.4cm}}

\newcounter{appendixc}
\newcounter{subappendixc}[appendixc]
\newcounter{subsubappendixc}[subappendixc]

\renewcommand{\appendix}[1] {\vspace*{0.6cm}
        \refstepcounter{appendixc}
        \setcounter{figure}{0}
        \setcounter{table}{0}
        \setcounter{equation}{0}
        \renewcommand{\thefigure}{\Alph{appendixc}.\arabic{figure}}
        \renewcommand{\thetable}{\Alph{appendixc}.\arabic{table}}
        \renewcommand{\theappendixc}{\Alph{appendixc}}

\renewcommand{\theequation}{\Alph{appendixc}.\arabic{equation}}
        \noindent{\bf Appendix \theappendixc #1}\par\vspace*{0.4cm}}

\def\abstracts#1{{

\centering{\begin{minipage}{12.2truecm}\footnotesize\baselineskip=12pt\noindent
	\centerline{\footnotesize ABSTRACT}\vspace*{0.3cm}
	\parindent=0pt #1
	\end{minipage}}\par}}

\renewenvironment{thebibliography}[1]
	{\begin{list}{\arabic{enumi}.}
	{\usecounter{enumi}\setlength{\parsep}{0pt}
\setlength{\leftmargin 1.25cm}{\rightmargin 0pt}
	 \setlength{\itemsep}{0pt} \settowidth
	{\labelwidth}{#1.}\sloppy}}{\end{list}}

\newcounter{itemlistc}
\newcounter{romanlistc}
\newcounter{alphlistc}
\newcounter{arabiclistc}

\newcommand{\fcaption}[1]{
        \refstepcounter{figure}
        \setbox\@tempboxa = \hbox{\footnotesize Fig.~\thefigure. #1}
        \ifdim \wd\@tempboxa > 6in
           {\begin{center}
        \parbox{6in}{\footnotesize\baselineskip=12pt Fig.~\thefigure.
#1}
            \end{center}}
        \else
             {\begin{center}
             {\footnotesize Fig.~\thefigure. #1}
              \end{center}}
        \fi}

\newcommand{\tcaption}[1]{
        \refstepcounter{table}
        \setbox\@tempboxa = \hbox{\footnotesize Table~\thetable. #1}
        \ifdim \wd\@tempboxa > 6in
           {\begin{center}
        \parbox{6in}{\footnotesize\baselineskip=12pt Table~\thetable.
#1}
            \end{center}}
        \else
             {\begin{center}
             {\footnotesize Table~\thetable. #1}
              \end{center}}
        \fi}

\def\@citex[#1]#2{\if@filesw\immediate\write\@auxout
	{\string\citation{#2}}\fi
\def\@citea{}\@cite{\@for\@citeb:=#2\do
	{\@citea\def\@citea{,}\@ifundefined
	{b@\@citeb}{{\bf ?}\@warning
	{Citation `\@citeb' on page \thepage \space undefined}}
	{\csname b@\@citeb\endcsname}}}{#1}}

\newif\if@cghi
\def\cite{\@cghitrue\@ifnextchar [{\@tempswatrue
	\@citex}{\@tempswafalse\@citex[]}}
\def\citelow{\@cghifalse\@ifnextchar [{\@tempswatrue
	\@citex}{\@tempswafalse\@citex[]}}
\def\@cite#1#2{{$\null^{#1}$\if@tempswa\typeout
	{IJCGA warning: optional citation argument
	ignored: `#2'} \fi}}

 1
 1
 1

\font\ninerm=cmr9

\begin{document}
\hfill CERN-TH/95-252

\hfil   \normalsize{\bf BOUNDARY EFFECTS IN STRING THEORY} \footnote{Talk
presented at the Strings '95 Conference, USC (March 1995).} \hfil
\baselineskip=16pt

\vspace*{0.6cm}
\centerline{\footnotesize MICHAEL B. GREEN{\footnote{ Permanent address: DAMTP,
Silver Street, Cambridge CB3 9EW, UK.}}}

\baselineskip=13pt
\centerline{\footnotesize\it TH Division, CERN}
\baselineskip=12pt
\centerline{\footnotesize\it CH-1211, Geneva 23, Switzerland}
\centerline{\footnotesize E-mail: m.b.green@damtp.cam.ac.uk}

\vspace*{0.9cm}
\abstracts{Some of the properties of string theory defined on
world-sheets with boundaries are reviewed.  Particular emphasis is
put on the possibility of identifying string configurations (\lq
D-instantons' and \lq D-branes') that give rise to stringy
non-perturbative effects.}

\vspace*{0.6cm}
\normalsize\baselineskip=15pt
\setcounter{footnote}{0}
\renewcommand{\thefootnote}{\alph{footnote}}
The perturbative expansion of closed-string theories is defined as a
sum over closed orientable Riemann surfaces.  Type 2a, 2b and
heterotic superstring theories are of this type while type 1
superstring theory has a perturbation expansion that involves a sum
over world-sheets with boundaries.  The presence of boundaries leads
to qualitatively new effects.  With conventional Neumann boundary
conditions there are open-string states as well as closed-string
states in the spectrum -- the boundaries represent the trajectories
of the string end-points.  Upon compactification to lower dimensions
$T$-duality changes this to a theory with Dirichlet boundary
conditions in the compactified dimensions$^{1,2}$.  The most drastic
modification arises in the  theory defined  by summing over
world-sheets with boundaries on which the string
space-time coordinates satisfy constant Dirichlet
conditions in {\it all} space-time dimensions -- in that case
the entire boundary is mapped to a point in the target space-time and
the
position of that point is then integrated, which restores
target-space
translation invariance  (ref.$2$ and references therein).    The
result is a
theory that describes closed strings which possess dynamical
point-like
substructure as is indicated by the fact that fixed-angle scattering
is power
behaved as a function of energy.  Recently an
interesting
variation of this  scheme has been suggested$^3$  (based on ref.$1$),
involving the idea of \lq D-instantons'.  As the name
suggests,  these are world-sheet configurations which correspond to
target
space-time \lq
events', giving rise to exponentially suppressed contributions to
scattering
amplitudes behaving as  $e^{-C/\kappa}$ (where $\kappa$ is the
closed-string
coupling constant that is determined by the dilaton expectation value
and $C$ is
a constant).   This is in accord with general observations based on
studying the convergence of string perturbation theory$^4$ that
suggest that
whereas instanton effects in  field theory typically behave as
$e^{-const./\kappa^2}$, analogous effects in closed-string theory
should behave
as
$e^{-C/\kappa}$.  Such  effects may be of significance in the
study of non-perturbative phenomena in superstring theory,
particularly in connection with the r\^ole  of BPS exact soliton
states in $U$ duality (see the talks by C.M. Hull and E. Witten at
this conference).

A novel feature of the sum over world-sheets with fixed Dirichlet
boundaries is the presence of momentum-independent divergences
arising from boundaries of moduli space at which open-string strips
degenerate$^2$.  It was argued$^3$ that these divergences cancel when
the sum over world-sheets is arranged with the combinatorics
appropriate to a $D$-instanton.   I will demonstrate how
a single D-instanton contribution
can be expressed as an exponential of an instanton \lq action'  and a
\lq gas'
of  such instantons can be defined.  This will clarify the
cancellation of the Dirichlet divergences (details of this argument
were published in ref.$5$ subsequent to the conference).  We will see
that the leading contribution to the free energy comes from free
D-instantons and is of
order $\kappa^{-1}$  but corrections due to
long-distance interactions between D-instantons will be seen to be of
order $\kappa^0$.

A general oriented string world-sheet has an arbitrary number of
boundaries and
handles.  With Dirichlet
boundary conditions each boundary is fixed at a space-time point,
$y_B^\mu$ (where $B$ labels
the
boundary), which is then integrated.
The boundaries of moduli space are of the following types.  Firstly,
there are the
usual
degenerations of cylindrical segments that correspond to the
propagation of
physical closed-string states:
\begin{itemize}
\item {(a)} Degeneration of handles.
\item{(b)} Degeneration of trivial homology cycles that divide a
world-sheet
into two disconnected pieces.
\end{itemize}

 \noindent In addition,  in the presence of boundaries there are
degenerations
of the following kinds:
\begin{itemize}
\item{(c)}  A boundary may shrink to zero length giving the
singularities
associated with closed-string scalar states coupling to the vacuum
through the
boundary.
\item{(d)} Degeneration of strips forming open-string loops.  If both
string
endpoints are fixed at the same target-space point this gives an
infinite
contribution.
\item{(e)} Degeneration of trivial open-string channels, in which the
world-sheet divides into two disconnected pieces.  In the case of
Dirichlet
conditions the intermediate open string necessarily has both
end-points fixed at
the same space-time point leading to another infinite contribution.
\end{itemize}

The BRST cohomology of the states of the open-string sector where the
string
end-points are fixed at $y_1^\mu$ and $y_2^\mu$ is isomorphic to that
of the
usual Neumann open string with momentum $p^\mu =  \Delta^\mu \equiv
y_2^\mu -
y_1^\mu$.  This may be viewed as a simple consequence of target-space
duality
and it means that the arbitrary diagram possesses a rich spectrum of
space-time
singularities, just as the usual loop diagrams possess a rich
momentum-space
singularity structure.  However, it is important that the
wave
functions of these states depend on the mean position, $y^\mu =
(y_1^\mu +
y_2^\mu)/2$, in addition to $\Delta^\mu$ -- this extra variable has
no analogue
for
the usual open strings.  The intermediate open string in the trivial
degeneration (e) has $\Delta=0$ so its cohomology is isomorphic to
that of the
usual Neumann open-string theory when $p^\mu=0$. There is only one
physical
state in this case, which is the level-one vector.  This is the
isolated
zero-momentum physical state with a constant wave function in the
usual theory.
 However, in the Dirichlet theory its wave function $\zeta^\mu(y)$
is an
arbitrary function that is physical without the need to impose any
constraints
on it  -- it is a target-space Lagrange multiplier field.  The
presence of this
as an  intermediate state in a string diagram leads to a divergence
(the
propagator for the level-one state is singular since a Lagrange
multiplier field
has no kinetic terms).  The vertex operator that describes the
coupling of this
level-one vector state to a boundary is given by
\begin{equation}
g\oint d\sigma_B \zeta\cdot \partial_n X(\sigma_B,\tau_B) = i g
\zeta^\mu
{\partial\over \partial y_B^\mu}\label{veccoup}
 \end{equation}
where  $n$ denotes the derivative normal to the boundary that is
fixed at the target-space position $y_B$ and $\tau_B$ is its
world-sheet position ($g$ is the open-string coupling
constant that is proportional to $\sqrt \kappa$).

There are two schemes for dealing with this level-one
divergence.  In
one of these the Lagrange multiplier field is eliminated by
integrating it,
thereby imposing a constraint  before the perturbation expansion of
the theory
is considered.  Some consequences of the presence of this constraint
were
discussed in  ref.$6$.  The other scheme$^3$ uses different
combinatorics
for the sum
over boundaries, giving rise to $D$-instantons.  In this
case  the
divergences due to the level-one open-string field are supposed to
cancel
between an
infinite number of diagrams as will be demonstrated below.

The construction of the partition function for a $D$-instanton gas
will involve the exponentiation of a particular combination of
contributions of connected world-sheets.  Expanding this exponential
therefore gives contributions from  world-sheets that are
disconnected (a concept that is not familiar in conventional
discussions of string perturbation theory).  We begin by obtaining an
equation for the contribution of the particular sum of connected
(orientable) world-sheets with Dirichlet boundaries that will later
be exponentiated.

Consider a connected
orientable world-sheet  with $p_i$ boundaries fixed at any one of a
finite number of points $y_i$ (where
$i=1, \dots, n$ and $p_i =0, \dots, \infty$).  The string free
energy, $f_{p_1,p_2, \dots,p_n}(y_1,y_2,\dots,y_n)$,
is given by the usual multi-dimensional integral over the moduli
space of the
surface which has a total of $\sum_i p_i$ boundaries.  We shall be
interested in the sum over surfaces with all possible numbers of
boundaries for a given value of $n$,
\begin{equation}
S^{(n)} = \sum_{p_1,p_2,\dots,p_n=0}^\infty {1\over p_1! p_2! \dots
p_n!}
f_{p_1,p_2, \dots,p_n}(y_1,y_2,\dots,y_n), \label{facts}
\end{equation}
where the explicit combinatorial factor accounts for the symmetry
under the
interchange of identical boundaries  -- boundaries that are fixed at
the same
space-time point.  The definition of $f_{p_1,\dots, p_n}$ implicitly
contains a
sum over handles and the term with all $p_i=0$ in Eq. (\ref{facts})
is just the
usual
closed-string free energy, $S^{(0)}$.   Both types of open-string
degenerations,
(d) and (e),  described earlier lead to divergences in Eq.
(\ref{facts}) due to
the
intermediate level-one open-string states and in each case the
coefficient of
the divergent term is proportional to the product of level-one vertex
operators
attached to the boundary at either end of the degenerating strip.  It
is
sufficient to consider single degenerations since the multiple
degenerations are
a subspace of these. The divergences of interest have the form
$\int_\epsilon^1
dq/q =\ln \epsilon$ where  $\epsilon$ is a world-sheet regulator.
Degenerations of type (e) divide the world-sheet into two factors, so
that if
the degenerating boundary is fixed at $y_1^\mu$ the singular term has
the form,
\begin{eqnarray}
&&\sum_{p_1,p_2,\dots, p_n;q_1,q_2,\dots,
q_n}{f_{p_1+q_1+1,p_2+q_2,
\dots,p_n+q_n} \over (p_1+q_1+1)! (p_2+q_2)! \dots (p_n+q_n)!}\sim
\nonumber\\
&&  \ln \epsilon \left( {\partial \over \partial y_1^\mu}
\sum_{p_1,p_2,\dots
,p_n} {f_{p_1,p_2,\dots,p_n} \over p_1! p_2!\dots p_n!} \right)
\left(
{\partial \over \partial y_{1\mu} } \sum_{q_1,q_2,\dots, q_n}
{f_{q_1,q_2,\dots,q_n}\over q_1! q_2! \dots q_n!}
\right),\label{edegen}
\end{eqnarray}
where the derivatives arise from two level-one  open-string vertex
operators
attached to two different boundaries.
A divergence also arises when an internal open-string with both ends
fixed at
the same point degenerates (this is a degeneration of type (d)).  In
this case
the coefficient of the divergence is proportional to two vertex
operators of the
level-one state attached to the same boundary giving,
\begin{eqnarray}
&& \sum_{p_1,p_2,\dots, p_n;q_1,q_2,\dots,
q_n}{f_{p_1+q_1+1,p_2+q_2,
\dots,p_n+q_n} \over (p_1+q_1+1)! (p_2+q_2)! \dots (p_n+q_n)!}\sim
\nonumber\\
 &&
\ln \epsilon {\partial^2\over \partial  y_1^2} \sum_{p_1+q_1,
p_2+q_2,\dots,
p_n+q_n} {f_{p_1+q_1, p_2+q_2,\dots, p_n+q_n} \over (p_1+q_1)!
(p_2+q_2)! \dots
(p_n+q_n)!} .
\label{ddegen}
\end{eqnarray}

Combining Eqs. (\ref{edegen}) and (\ref{ddegen}) and taking a
derivative of the free
energy with
respect to $\epsilon$  extracts the dependence on the divergent
degenerations,
\begin{equation}
 \epsilon {\partial \over \partial \epsilon}  S^{(n)} = \left(
{\partial \over
\partial y_1^\mu} S^{(n)} \right) \left( {\partial \over \partial
y_{1\mu}
}S^{(n)} \right)   + {\partial^2\over \partial  y_1^2} S^{(n)}
{}.
\label{degenone}
\end{equation}

The rules for constructing the string partition function in the
presence of a
single D-instanton$^3$ were presented as an infinite series of terms
as follows.   Firstly,
sum over world-sheets with insertions of any number of handles and
Dirichlet
boundaries that are all at the {\it same point} in the target space,
$y_1^\mu$,
which is to be integrated over.   The sum is now taken to include
{\it
disconnected} world-sheets although these do not appear to be
disconnected from
the point of view of the target space since the boundaries all touch
the same
point.  A suitable factor  is to be included to take account
of
symmetry under the interchange of identical disconnected
world-sheets.  This series of terms can be expressed as an
exponential in terms of the connected world-sheets that were
introduced earlier giving the
one D-instanton partition function,
\begin{equation}
Z^{(1)} = \int d^Dy e^{S^{(1)}(y)}.
\label{partfuh}
\end{equation}
The exponent $S^{(1)}(y) = S^{(0)} + \sum_{p=1}^\infty f_p(y)/p!$ is
now
interpreted as the one
D-instanton \lq action'
that is given by the $n=1$ term in Eq. (\ref{facts}).  Scattering
amplitudes may
be
generated from this expression if   $S^{(1)}$ is taken to be a
functional of the
background fields.

The requirement of consistent clustering properties in the target
space (as well as on the world-sheet) motivates
the following generalization that includes the sum over an arbitrary
number of
D-instantons (and which should presumably be equivalent to the rather
schematic
generalization motivated by  duality$^3$).  This
involves summing over insertions of boundaries at any number of
positions,
$y_i$, that are to be integrated. The partition function is given in
the
language of a conventional instanton gas by the expression,
\begin{equation}
Z = \sum_n {1\over n!}\left( \prod_{i=1}^n d^D y_i^\mu \right) e^{
S^{(n)}(y_1, \dots, y_n)},
\label{partsuma}
\end{equation}
where $S^{(n)}$ is given by Eq. (\ref{facts}) and is now interpreted
as the
action for $n$
interacting D-instantons.  Recall that  $S^{(n)}$ is defined by a
functional
integral over connected world-sheets and includes a term with no
boundaries
which is equal to $S^{(0)}$, the usual closed-string free energy.
The
expression for $Z$ therefore has the form,
\begin{equation}
Z= e^{S^{(0)}}\sum_n {1\over n!}\left( \prod_{i=1}^n d^D y_i^\mu
\right) e^{
S^{(n)\prime}(y_1, \dots, y_n)},
\label{partsumb}
\end{equation}
where $S^{(n)\prime}$ is defined to be $S^{(n)}$ with the
zero-boundary term
missing.  In the general term in the sum any  boundary may be located
at any one
of the $n$ target-space positions, $y_i^\mu$, which are analogous to
the
collective
coordinates describing the positions of instantons in quantum field
theory.   It
is convenient to decompose $S^{(n) \prime}$ into those terms in which
all
boundaries are fixed at the same point (the free D-instanton terms),
those at which the boundaries are fixed at two different points
(two-body D-instanton interaction terms), those involving
three points (three-body D-instanton interactions), and so on,
\begin{equation}
S^{(n)\prime}(y_1, \dots, y_n) = \sum_{i=1}^n R_1 (y_i) +
\sum_{i\ne j}^n R_2
(y_i,y_j)+ \sum_{i\ne j\ne k}^n R_3 (y_i,y_j,y_k) + \dots
.\label{instact}
\end{equation}
It is important that the definition
of
$S^{(n)\prime}$ includes terms in which any subset of the $p_i$ are
zero --
these are terms that also contribute to the definition of
$S^{(m)\prime}$ with
$m<n$.
\begin{figure}
\begin{center}
\leavevmode
\epsfbox{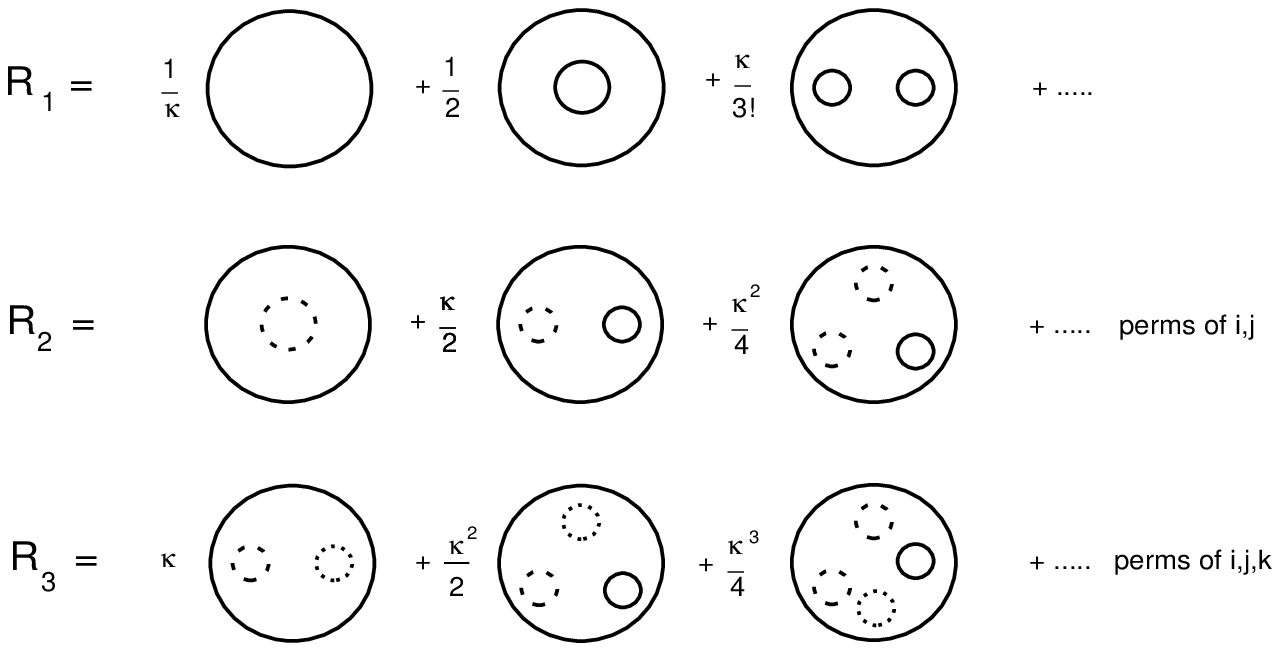}
\end{center}
\caption[]{ \footnotesize Contributions to the $n$ D-instanton action
from the
first few
powers of $\kappa$.  a)  Diagrams with boundaries fixed at a single
space-time
point contribute to the free action.  b)  Diagrams with boundaries
fixed at two
points (indicated by the full and dashed boundaries) contribute to
the
two-instanton interaction.  c)  Diagrams contributing to the
three-instanton
interaction (at points indicated by full, dashed and dotted
boundaries).  The
coefficients explicitly show the combinatorical factors that arise
from symmetry
under the interchange of identical boundaries.
}
\end{figure}


The free term in Eq. (\ref{instact}), $R_1(y_i)\equiv
S^{(1)\prime}(y_i)$, is
simply
given by the sum over connected orientable world-sheets of arbitrary
topology
with all boundaries fixed at a single point, $y_i$, illustrated in
fig.1(a).  It is independent of $y_i$ (by translational invariance)
and
has the form,
\begin{equation}
R_1=  -{C\over \kappa} + \ln D + O(\kappa),
\label{leadterm}
\end{equation}
and thus $\sum_i R_1(y_i) = n R_1$.   The term in this series with
constant
coefficient, $C$, is determined by  functional integration over the
disk.
Explicit calculations determine $C=  2^8 \pi^{25/2} \alpha^{\prime
6}$$^{7,8}$.
It is somewhat remarkable that $C$ is a finite (positive) constant
with a value
that is consistent with the non-vanishing value of the disk with a
zero-momentum
dilaton attached.  Naively $C$ would be expected to vanish since it
should be
proportional to the inverse of the volume of the conformal Killing
group,
$SL(2,R)$ (which is infinite) but that would not be consistent with
the disk
with a zero-momentum dilaton insertion.   The $\zeta$ function
regularization$^7$ and the proper-distance regularization$^8$ lead
to the
subtraction of an infinite constant from the volume of the Killing
group, giving
the finite positive renormalized value of $C$.  The
$\kappa$-independent
constant $D$ in
Eq. (\ref{leadterm}) comes from the world-sheet annulus with both
boundaries at
$y^\mu$.

Equation (\ref{leadterm}) leads to $e^{-C/\kappa}$ contributions to
the
partition
function and to scattering amplitudes.  This has the qualitative form
expected
for non-perturbative effects in string theory on the basis of  matrix
models and
from the analysis of the rate of divergence of closed-string
perturbation theory$^4$.  It is to be contrasted with a
characteristic feature
of
non-perturbative effects (such as instantons and solitons) in field
theory,
which behave as $e^{-const./\kappa^2}$.  This distinction between the
non-perturbative behaviour of quantum field theory and that expected
in
closed-string theory seems likely to be of great significance (some
possible
consequences are described in ref.$9$).

The two-instanton interactions are given by the series of terms in
$R_2$  shown
in fig.1(b).  The diagrams contributing to $R_2$ are those in
which at least one boundary is fixed at either of the two space-time
points.  The leading terms in
this series have the form
\begin{equation}
R_2(y_i,y_j) =  f_{1,1}(y_i,y_j)   + {\kappa\over 2} \left(
f_{2,1}(y_i,y_j)
 +  f_{1,2}(y_i,y_j) \right)+ O(\kappa^2),
\label{twobody}
\end{equation}
where $f_{p_i,p_j }(y_i,y_j )$ indicates a term with all $p_r=0$
apart from
$p_i$ and $p_j$.   The two-boundary term  is given by the expression
\begin{equation}
 f_{1,1}(y_i,y_j)  = c \int_0^\infty d\tau e^{-\Delta_{ij}^2 /\tau}
e^{2\tau}
\prod_{n=1}^\infty (1-e^{-2\tau})^{-24},
\label{twodiff}
\end{equation}
where $c$ is a constant and $\Delta_{ij} = y_2-y_1$.
This diverges at the endpoint $\tau\to \infty$ due to the presence of
a
closed-string tachyon state.  This is a familiar problem of the
bosonic theory
which we shall bypass by tranforming to momentum space and declaring
that at low
momenta (or large distance) only the massless dilaton singularity
survives so
that in the  long-distance limit $\Delta^2 \to \infty$
\begin{equation}
f_{1,1}(y_i,y_j) \sim   |y_i - y_j|^{ 2-D} .
\label{longdist}
\end{equation}
This coulomb-like behaviour due to dilaton exchange is analogous to
the
long-distance force between two classical instantons (magnetic
monopoles) in the
three-dimensional euclidean Georgi--Glashow model.  However, unlike
the case of
magnetic monopoles the interaction term, Eq. (\ref{longdist})  is not
of the
same order in
$\kappa$ as the leading term,  $S^{(1)\prime}$.

It is straightforward to show that the partition function defined by
Eq. (\ref{partsumb}) does not have the divergences arising from the
level-one
open-string
vector state.  The term in the partition function coming from $n$
D-instantons
has
a dependence on $\epsilon$ that can be written by expanding the
exponent to
first order in $\ln \epsilon$ using Eq. (\ref{degenone}), giving,
\begin{equation}
\ln \epsilon \int \prod_{i=1}^n d^Dy_i^\mu \left\{\left(
{\partial
\over \partial y_1^\mu} S^{(n)} \right) \left( {\partial \over
\partial y_{1\mu}
}S^{(n)} \right)   + {\partial^2\over \partial  y_1^2}
S^{(n)}\right\}  e^{
S^{(n)}}
 = 0.
\label{finitr}
\end{equation}
The fact that the expression vanishes makes use of an integration by
parts of
the second term.

The cancellation is illustrated in an example in fig.~3.  This shows
contributions to a particular divergence coming from the sum of (a) a
planar
connected world-sheet, (b) a planar disconnected world-sheet and (c)
a
non-planar disconnected world-sheet.    The sum of these
contributions may be
written symbolically as
\begin{equation}
(a)+ (b) + (c) \ = \  \ln \epsilon \int d^Dy_1^\mu{\partial^2\over
\partial
y_1^2}\left( f_{1,1}(y_1,y_2)
f_{2,1}(y_1,y_2)\right),
\label{abcdegen}
\end{equation}
which vanishes after integration over $y_1$ (assuming suitable
boundary
conditions).
The cancellation of divergences evidently involves a conspiracy
between terms
with different numbers of boundaries and handles.  Therefore it is
only possible
when the
boundary weight has a specific value -- it is not possible to add
Chan--Paton
factors to the boundaries as is usually the case in open string
theories.
It is disturbing that the cancellation of the divergences requires an
integration by parts which looks nonlocal since the D-instanton gas
is supposed
to
satisfy clustering properties that express the locality of the
theory.  However,
(at least in flat space) the potentially dangerous surface terms that
arise are
suppressed since they involve interactions between boundaries fixed
at widely
separated points.
\begin{figure}
\begin{center}
\leavevmode
\epsfbox{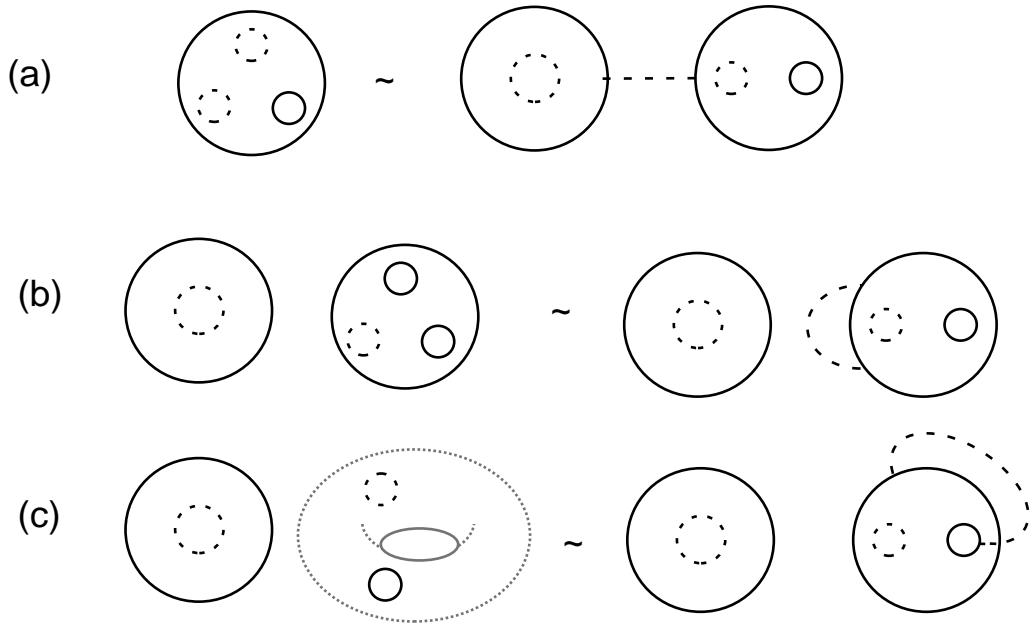}
\end{center}
\caption[]{\footnotesize An example of the cancelation of a
divergence that
requires
world-sheets with handles. a) One of the divergent degenerations of a
world-sheet with two boundaries at $y_1$ (full lines) and two at
$y_2$ (dashed
lines).  b)  A degeneration of a disconnected planar  world-sheet
that, after
integration over $y_1$,  contributes to the same divergence as in a).
 c)   A
degeneration on a disconnected world-sheet with a handle that gives a
divergence
that adds to  the divergences in b) and c) to give a total $y_1$
derivative.}
\end{figure}


On-shell scattering amplitudes may be defined in the usual manner by
considering
fluctuations in the background fields, resulting in closed-string
vertex
operators coupled to the world-sheets.  The connected scattering
amplitudes are
obtained
from $\ln Z$, where the word \lq connected' here refers to the target
space.  It is a important that the expression for $\ln Z$ only
generates diagrams that are connected in the target space -- in other
words that disconnected world-sheets only arise when the disconnected
components have at least one boundary fixed at a
common target-space point.  This follows from combinatorics that are
similar to the usual field theory combinatoric arguments that show
that the perturbative expansion of $\ln Z$ generates connected
Feynman diagrams.    The terms
that are disconnected in the target space (terms in which vertex
operators are attached to disconnected world-sheet components that do
not have any boundary
fixed at a common point) can be shown to cancel out of the expansion
of $\ln Z$.   The presence of Dirichlet boundaries gives terms of
order $e^{-C/\kappa}$ in scattering amplitudes that have
qualitatively different behaviour from the usual behaviour.  For
example, there is a contribution to the four-tachyon amplitude with
each vertex operator attached to a separate component of the
world-sheet -- this is a $\phi^4$ contact interaction.  The
lowest-order contribution to the scattering amplitude with four
physical gravitons (which comes from a diagram with a pair of vertex
operators attached to two disks with boundaries at the same point in
the target space) falls as a power of the energy$^5$.

The inconsistencies of the critical bosonic string  make it difficult
to interpret this theory in more detail. In particular, it is not at
all
clear in
what sense these ideas  make contact with more conventional instanton
ideas, such as those that arise in matrix models.   It would be of
interest
to
study similar boundary effects in two-dimensional bosonic string
theories  and
compare them with other descriptions of instantons in such theories
(such
as  ref.$10$). One peculiarity, at least in the
bosonic theory,
is that there are no anti D-instantons.

The absence of tachyons in  superstring theories  suggests that this
may be an arena where a more
consistent discussion could be given.  The very recent interest in
the possible connection between type 1 superstrings and the heterotic
string (E.  Witten at this conference and  refs.$11,12$ subsequently)
gives added impetus to the study of the effect of unusual boundary
conditions in superstring theory  (previous work on the construction
of type 1 theories in various dimensions is described in  ref.$13$
and references therein).   In the type 2b theory there are two types
of Dirichlet boundaries, each preserving one half of the space-time
supersymmetry$^{14}$ -- these may be thought of as self-dual and anti
self-dual.
A gas of self-dual (or anti self-dual) super $D$-instantons is
non-interacting just as in the case of self-dual monopoles.  This
follows because the  multi self-dual D-instanton action is given by a
sum over connected world-sheets with any number of boundaries of the
same type that are fixed at any of n points.  This configuration has
world-sheet supersymmetry and the integration over fermion zero modes
kills all diagrams with more than one boundary.  Therefore for n
self-dual super D-instantons the free energy is simply n times the
free energy of a single free D-instanton -- n  times the disk
diagram, which is again a finite constant, given by the non-zero
coupling of
the dilaton
to the disk with one particular spin structure of the world-sheet
fermions.

There are also soliton-like  \lq D-brane' configurations$^3$ whose
r\^ole in the context of superstrings has not yet been illuminated.
However these  might provide solitonic states that are needed if the
suggested non-perturbative equivalence of the type 1 and heterotic
theories is correct.

\end{document}